# Toward the regularization of E value from BLAST similarity search into a dissimilarity measure as distance function, and the metrication of protein sequence space*

Boryeu Mao

Independent Unaffiliated Researcher
Redmond, Washington 98052, United States

boryeu.mao@gmail.com

Running Title: Toward an E value distance function and metrication of protein sequence space

## Abstract

Sequence matching algorithms such as BLAST and FASTA have been widely used in searching for evolutionary origin and biological functions of newly discovered nucleic acid and protein sequences. As parts of these search tools, alignment scores and E values are useful indicators of the quality of search results (and the relevance of the matches) from querying a database of annotated sequences, whereby a high alignment score (and inversely a low E value) reflects significant similarity between the query and the subject (target) sequences. For cross-comparison of results from sufficiently different queries however, the interpretation of alignment score as a similarity measure and E value a dissimilarity measure becomes somewhat nuanced, and prompts herein a judicious distinction of different types of similarity. Via a simulated formulation, we show that an adjustment of E value to account for self-matching of query and subject sequences corrects for certain ostensibly anomalous similarity comparisons, resulting in 'regularized' dissimilarity and similarity measures that would be more appropriate for cross-comparisons, as well as database applications, such as all-on-all sequence alignment or selection of diverse subsets. In actual practice, the 'regularization' of E value dissimilarity improves clustering and subset selection. While both E value and the 'regularized' E value share two of the four axiomatic properties of a metric space, positivity and symmetry, the latter E value further becomes reflexive and meets the condition of triangle inequality, the remaining two axioms, thus itself an appropriate distance function for metricating protein sequence space.

## Introduction

For nascent nucleic acid and protein sequences, often the first step in the identification of the evolutionary origin and the biological function is the search for similar sequences in annotated bioinformatics databases. Sequence matching tools such as BLAST [1-4] evaluate molecular similarity for proteins and nucleic acids with algorithms for aligning sequences and computing alignment scores [4, 5] according to the extent of amino acid residue or nucleic acid base matches and mismatches along the length of the sequences, as well as any gaps that may help improve the overall alignment. Calculated from the alignment score for a query-subject sequence pair, an expected value, or E value [2,5-7], is a statistical estimate of the expected number of chance matches with better alignment, a useful indicator of the significance and relative quality of search results from screening the given query sequence against a relevant, annotated bioinformatics database. The exponential relationship between the alignment score `S` and the E value `Eval`, in the functional form of `Eval`~ e^(−`S`) [2,5], is an inverse relationship in the general form of an exponential decay function for transformation between a similarity measure and its dissimilarity counterpart, and vice versa [8,9]. The comparison of `S` and `Eval` for an example set of five sequences, listed in Table I(a) and derived from protein domain ***d1dlwa_*** in the hierarchical protein structure database `SCOP` [10,11], are shown in Table I(b). For query sequence `seq.1`, the alignment score decreases as the subject (target) sequence becomes less similar (progressively shorter), from `seq.z` to `seq.b`, and to `seq.az`, with concomitant increases of E value. Relative to the `seq.1` series, the alignment score of the subject/query pair of `seq.az/a` is higher than `seq.az/1` as expected (since `seq.a` is closer to `seq.az` than `seq.1` is), but somewhat unexpectedly falls short of `seq.b/1`, despite the fact that with only one substitution the `seq.az/a` pair might've been considered to be more similar than `seq.b/1`, for which there is a gap and a length deficit of 16 residues in the global alignment. This apparent contradiction, that a higher similarity is reflected in a higher score in one instance (`seq.b/1` over `az/1` with a score of 480.0 over 361.0) but paradoxically in a lower score in another (`seq.az/a` over `b/1` and yet with a score of 407.0 below 480.0), is partly semantic, and largely resolvable with the recognition of two distinctive types of similarity: (i) the alignment score-based similarity which is strongly influenced by the extent of pairwise matched positions between query and subject, and (ii) a 'regularized' form of similarity unified across queries, with which `seq.az/a` would otherwise score higher than `seq.b/1`. A corollary of the above is that, while the alignment-score similarity (and the inverse dissimilarity) is completely satisfactory for evaluating matches from a single, or closely related queries, a suitably defined 'regularized' similarity measure may be more appropriate in order for matches from sufficiently different queries to be directly cross-compared more efficaciously. This putative 'regularized' similarity and the related dissimilarity measure may be expected to be equally suitable for comparing single query matches as well as those from disparate queries such as the all-on-all sequence matching within a database for discovering homologous sequences, detecting gene families, constructing phylogenetic trees [12,13], clustering sequences for diversity selections [14], or constructing protein families [15,16].

Table I

In the Results and Discussion section, investigations of the five protein sequences in Table I(a) comprise Case Study #1 for illustrating the ostensible anomaly, and for motivating the regularization of E value for resolving the anomaly. For facilitating calculations on sequence sets in the Case Studies, a stand-alone system of procedures was prepared for simulating the E value computation in BLAST searches, which is described in the Methods section with details on parameters for the calculation of alignment scores and E values.  Formal properties of the sequence length-adjusted, 'regularized' E value are then examined in Case Studies #2 - #4, on sequences of `SCOP` structure domains, and their parametric dependency noted as appropriate.

Also in the Methods section, relevant formulae, expressions and systems information are collected and grouped into subsections A to H.  Key items are labeled in bold.

## Results and Discussion

### **A. Case Study #1, five sequences derived from `SCOP` domain *`d1dlwa_`* (Table I(a))**

Table II

`SCOP` domain ***`d1dlwa_`*** is fetched with `Biopython` package `Bio.SCOP`, class and method `Scop.getDomains()[1]` from the database (see Methods section). Alignment score `S` and E value `Eval`, calculated respectively from expressions `M.1` and `M.2` for four relevant pairs (column 1 of Table II, rows 1-4), are shown in columns 2 and 3. Against the series of pairwise alignments for query `seq.1` (rows 1-3), the alignment score `S` for the subject/query pair `seq.az/a` (row 4, same length with one substitution) is higher than that of `seq.az/1` (row 3, length deficit of 36 residues), as expected. The alignment score of 407.0, as a similarity measure for the `seq.az/a` pair, is lower than 480.0 for the `seq.b/1` pair however, contrary to the expectation that the single substitution in the `seq.az/a` pair presumably should imply a higher degree of similarity (and thus a higher score) than the 16-residue length deficit in the `seq.b/1` pair.

Rather than a completely different similarity measure, either within or possibly without the current E value framework [17-19], the apparent contradiction (or at least an inconsistency, semantic or otherwise) may be resolved instead with the recognition, and the reconciliation and regularization, of two distinctive types of similarity. First, given that the alignment score accounts for pairwise similarity by counting matched positions, and penalizing mismatches with substitution scores and gap costs, a longer sequence would generate a higher score inherently from the more numerous positions to be examined in the matching. This type of similarity derived from alignment score might be provisorily qualified as a 'bits' similarity (and the corresponding 'bits' dissimilarity, headings of columns 2 and 3,

Table II), in somewhat the same way that a 'bit' score was defined from alignment score [3,20]. Every sequence as single query largely establishes an inherent 'reference frame' for computing the alignment scores with subject sequences, a reference frame within which alignment scores can be directly compared and the similarity or dissimilarity is completely well-defined. The 'bits' qualification only becomes necessary and significant when comparing results from different query sequences, in reference frames with different baselines and scales. Therefore a second, ‘regularized’ similarity is to be called upon for suitably reconciling and standardizing different reference frames from different query sequences, a similarity measure with which the sequence pair seq.az/a would appropriately score higher than seq.b/1, for instance.

Consistent with the notion of 'bits' similarity, the self-matching alignment score also shows ostensible length dependency (rows 5-7, Table II): seq.1 is more self-similar than either seq.b or seq.a because there are more residues in the sequence and consequently more positions, or 'bits', to contribute to the alignment score. This length dependency motivated the formulation of a 'base' E value $\mathrm{Eval}_{b}$, in the form of the geometric mean of self-matching alignment scores of participating sequence pair subject and query (Expr M.5), to be applied as a ‘regularization’ of E value Eval to $\hat{\mathrm{E}}\mathrm{val}$ (Expr M.6)[1] ideally suited for comparison across different queries. $\hat{\mathrm{E}}\mathrm{val}$ values are shown in column 5, and in column 6 the inverse, i.e. the corresponding similarity measure $\hat{S}$ (Expr M.8). Note that the ‘regularized’ similarity $\hat{S}$ for seq.az/a is now higher than that for seq.b/1 as desired. Since the alignment score $S(\overline{s},\overline{q})$ is always smaller than $S(\overline{s},\overline{s})$ or $S(\overline{q},\overline{q})$ for any subject s and

*[1] Similarity and dissimilarity measures derived from $\hat{\mathrm{E}}\mathrm{val}$ are referred to as ‘regularized’ similarity or dissimilarity hence forth, to be understood as a regular or ordinary similarity/dissimilarity measure.*

query $\mathtt{q}$ due to mismatches and/or gaps in their respective amino acid sequences $\overline{\mathtt{s}}$ and $\overline{\mathtt{q}}$,

$\mathtt{Eval}(\overline{\mathtt{s}},\overline{\mathtt{q}})$ would be larger than $\mathtt{Eval}_{\mathtt{b}}(\overline{\mathtt{s}},\overline{\mathtt{q}})$, and from Expr $\mathtt{M.6}$

$$\hat{\mathtt{E}}\mathtt{val}(\overline{\mathtt{s}},\overline{\mathtt{q}}) := \ln(\mathtt{Eval}(\overline{\mathtt{s}},\overline{\mathtt{q}}) / \mathtt{Eval}_{\mathtt{b}}(\overline{\mathtt{s}},\overline{\mathtt{q}})) > \ln(1)$$

and thus,

$$\hat{\mathtt{E}}\mathtt{val}(\overline{\mathtt{s}},\overline{\mathtt{q}}) > 0 \qquad \mathtt{R.1}$$

Notably, in contrast to $\mathtt{Eval}$'s, $\hat{\mathtt{E}}\mathtt{val}(\overline{\mathtt{z}},\overline{\mathtt{1}})$ and $\hat{\mathtt{E}}\mathtt{val}(\overline{\mathtt{az}},\overline{\mathtt{a}})$ are both 1.602 (column 5, Table II), each pair being a single threonine-to-glutamine substitution (Table I(a)). For the three self-matching pairs in rows 5-7, Table II, the subject sequence ***is*** the query sequence; therefore $\overline{\mathtt{s}} = \overline{\mathtt{q}}$, and from Expr $\mathtt{M.5}$,

$$\mathtt{Eval}_{\mathtt{b}}(\overline{\mathtt{q}},\overline{\mathtt{q}}) = \mathtt{Eval}(\overline{\mathtt{q}},\overline{\mathtt{q}})$$

and from Expr $\mathtt{M.6}$,

$$\hat{\mathtt{E}}\mathtt{val}(\overline{\mathtt{q}},\overline{\mathtt{q}}) = 0 \qquad \mathtt{R.2}$$

whereas self-matching 'bits' dissimilarity $\mathtt{Eval}(\overline{\mathtt{q}},\overline{\mathtt{q}})$ is generally greater than 0.0, as shown in Table II, column 5 vs column 3, rows 5-7. By virtue of $\mathtt{R.2}$, $\hat{\mathtt{E}}\mathtt{val}$ is a proper ***distance*** function that by definition must be null for any sequence to itself, hence named 'distance E value' in subsection <u>D</u> in <u>Methods</u>.

Since the query and subject sequences are of different length in general, $\mathtt{Eval}(\overline{\mathtt{s}},\overline{\mathtt{q}}) \neq \mathtt{Eval}(\overline{\mathtt{q}},\overline{\mathtt{s}})$ (Expr $\mathtt{M.2}$), and thus $\hat{\mathtt{E}}\mathtt{val}(\overline{\mathtt{s}},\overline{\mathtt{q}}) \neq \hat{\mathtt{E}}\mathtt{val}(\overline{\mathtt{q}},\overline{\mathtt{s}})$. Following Brenner et al. [14], the smaller of the two $\mathtt{Eval}$ values is assigned to $\mathtt{Eval}_{2}$ for the subject/query pair (Expr

Figure 1

$\mathtt{M.4}$), and similarly for $\hat{\mathtt{E}}\mathtt{val}_2$ (Expr $\mathtt{M.7}$). The difference between the two asymmetric $\hat{\mathtt{E}}\mathtt{val}$'s is generally small in magnitude[2], for example $\mathtt{seq.b/1}$ and $\mathtt{seq.az/1}$, rows 2,3, column 5 in Table II, and others in symbols in green in Figure 1(a).

Both $\mathtt{Eval}_2$ and $\hat{\mathtt{E}}\mathtt{val}_2$ are symmetrical, from $\mathtt{M.4}$, $\mathtt{M.7}$,

$$\mathtt{Eval}_2(\overline{\mathtt{s}},\overline{\mathtt{q}}) = \mathtt{Eval}_2(\overline{\mathtt{q}},\overline{\mathtt{s}}) \qquad \mathtt{R.3}$$

$$\hat{\mathtt{E}}\mathtt{val}_2(\overline{\mathtt{s}},\overline{\mathtt{q}}) = \hat{\mathtt{E}}\mathtt{val}_2(\overline{\mathtt{q}},\overline{\mathtt{s}}) \qquad \mathtt{R.3a}$$

$\hat{\mathtt{E}}\mathtt{val}_2$'s (and $\hat{\mathtt{S}}_2$'s) are shown in bold in Table II. Note relationships $\mathtt{R.1}$ and $\mathtt{R.2}$ also hold for $\hat{\mathtt{E}}\mathtt{val}_2$:

$$\hat{\mathtt{E}}\mathtt{val}_2(\overline{\mathtt{s}},\overline{\mathtt{q}}) > 0 \qquad \mathtt{R.1a}$$

$$\hat{\mathtt{E}}\mathtt{val}_2(\overline{\mathtt{q}},\overline{\mathtt{q}}) = 0 \qquad \mathtt{R.2a}$$

The relationships for $\mathtt{Eval}_2$, $\mathtt{Eval}_\mathtt{b}$, and $\hat{\mathtt{E}}\mathtt{val}_2$ among various subject/query pairs in Table II are shown in Figure 1(a). The comparisons of the clustering dendrograms for $\mathtt{Eval}_2$ and $\hat{\mathtt{E}}\mathtt{val}_2$ are shown in Figure 1(b) and 1(c) respectively, summarizing numerical results in Table II.

---

***2*** *The difference between* $\hat{\mathtt{E}}\mathtt{val}(\overline{\mathtt{s}},\overline{\mathtt{q}})$ *and* $\hat{\mathtt{E}}\mathtt{val}(\overline{\mathtt{q}},\overline{\mathtt{s}})$ *is the difference between the logarithms of the subject and query sequence lengths according to Expr* $\mathtt{M.7}$, $\mathtt{M.6}$, *and* $\mathtt{M.2}$. *In Figure 1(a), the differences are no larger than 1.1%. The upper bound for values calculated from protein domains in* SCOP *database is about 1.3%, with an average of about 0.16%.*

Lastly, it can be readily verified that for rows 4,8,9, Table II, triangular inequality, Expr M.10, holds for ‘regularized’ dissimilarity Êval, but not 'bits' dissimilarity Eval:

$$\mathrm{Eval}(\overline{az},\overline{a}) + \mathrm{Eval}(\overline{z},\overline{a}) - \mathrm{Eval}(\overline{z},\overline{az}) < 0 \qquad \mathrm{R.4}$$

$$\hat{\mathrm{E}}\mathrm{val}(\overline{az},\overline{a}) + \hat{\mathrm{E}}\mathrm{val}(\overline{z},\overline{a}) - \hat{\mathrm{E}}\mathrm{val}(\overline{z},\overline{az}) \geq 0 \qquad \mathrm{R.4a}$$

To summarize, the alignment score of 407.0 is a proper 'bits' similarity measure of the seq.az/a pair. It is ***not*** an under-estimate *per se*, but ostensibly becomes one only as a substitute similarity measure for comparing results from multiple queries when no distinction is made between 'bits' similarity and ‘regularized’ similarity. Due to the inflated dissimilarity Eval relative to Êval, the seq.az/a pair would be mistakenly missed, and thus a potential false negative (PFN), in a near neighbor search. As shown in Table II, ‘regularized’ dissimilarity measure Êval readily recovers false negatives (column 5 vs. column 3) that could potentially impact the outcome of operations such as the construction of protein families [15,16]. Among the family of Êval functions parameterized in substitution matrix and gap scores (and corresponding $\lambda$ and K, M.7, M.6, M.2), important characteristics on PFN's and the triangle inequality relations are preserved (noted and shown as examples in Table II).

**B. Case Study #2, first 40 domains in SCOP**

The first 40 domains in SCOP are fetched from the database (version 2.08, updated on 2023-01-06) with Biopython package Bio.SCOP, class and method Scop.getDomains()[0:40]. In Table III(a), sequences of six of the domains[3] belong in three groups each of degeneracy of 2: `seq.1` and `2`, `seq.3` and `4`, and `seq.18` and `20`. The 'bits' similarity and dissimilarity of sequences within a group show length dependency among different groups (Table III(b), columns 2 and 3), same as the self-matching values in Case Study #1 (Table II, rows 5-7). In particular, the values for the pair `seq.1/2` mirror those for the self-matching of `seq.1` (row 5, Table II). Here `dom.1` and `2` are two distinctive 3D protein structure domains from separate X-ray diffraction experiments but share the same amino acid sequence, `seq.1` = `seq.2`, in contrast to the self-matching of a single sequence `seq.1` of domain `dom.1` in Table II; by the same token, $\hat{E}\mathtt{val}_2(\bar{1},\bar{2})$ is a distance of 0 between two domains coincidentally of identical sequences, whereas $\hat{E}\mathtt{val}_2(\bar{1},\bar{1})$ is the self-distance of `seq.1` which would be 0 by necessity. If the notion of 'bits' similarity were to be extended to 'bits' identity, then `seq.18/20` would be 'more identical' (in the 'bits' sense) than `seq.3/4` or `seq.1/2`, whereas the ‘regularized’ identity (arising from ‘regularized’ dissimilarity and similarity) would be universal for all domain groups with degenerate sequences (columns 4, 5, Table III(b)). For identity as a limiting case and a form of exact similarity, the ‘regularization’ perhaps would prove to be more consequential than a mere semantics exercise.

Table III

Figure 2 shows the clustering of the set of 40 domains according to either $\mathtt{Eval}_2$ or $\hat{E}\mathtt{val}_2$ as the dissimilarity measure. In Figure 2(b), degenerate sequence groups other than the three doubly-

Figure 2

---

***3*** *Protein domains from* `SCOP` *database (Table III) are denoted as* `dom.d`*, where* `d=1,2,...`*, and their amino acid sequences are denoted as* `seq.d` *as a short hand. Formally* `dom.d` *are protein structure domains with 3D conformation data, and* `seq.d` *are their amino acid sequences entered into the calculation of alignment scores and E values as query or subject.*

degenerate groups in Table III are readily recognizable by the merge height[4] of 0: one group of degeneracy of 3, three groups of degeneracy of 4, and one group of degeneracy of 8.

In addition to the identification of sequence-degenerate groups, the `Eval`-`Êval` relationship (**viz.** a plot similar to that colored in blue in Figure 1(a)) shows that 87.7% of the sequence pairs of the set of 40 `SCOP` domains are potential false negatives. And, for `Eval`'s, of 83.5% of the triplets the triangle inequality is violated.

### C. Case Study #3, first 180 domains of the ASTRAL domain subset @E value of 1e-50 in `SCOP`

Figure 3

The ASTRAL compendium of the `SCOP` database provides protein structure domain subsets[5] according to E value thresholds ranging from 1.0e−50 to 1.0e+1 [14]. From the subset file `astral-scopedom-seqres-sel-gs-e100m-verbose-e-50-2.08.txt`, at the lowest threshold E value of 1.0e−50, the Stable domain identifiers (`sid`) of the first 180 domains are extracted and the sequences fetched with `Astral.getSeq(Scop.getDomainBySid(`*`sid`*`))` individually. Figure 3 shows the clustering of domain sequences according to $\mathrm{Eval}_2$ in Figure 3(a) or $\hat{\mathrm{E}}\mathrm{val}_2$ in Figure 3(b). At the E value threshold of 1.0e−50, the 11 groups in Figure 3(a) correctly reconstruct the same clusters in the ASTRAL subset file. In Figure 3(b), re-clustering with $\hat{\mathrm{E}}\mathrm{val}_2$ shows that the 10 clusters in the branch colored orange are now reduced to 5; of the 5 reductions, 3 are

---

**4** *Merge heights are dissimilarity levels at which leaves and branches merge in a clustering tree, numerically marked on the Y-axis of dendrogram plots in Figures 1-3.*

**5** *Rather than protein structure domains, the unit of classification of* `SCOP` *database, it is possible, and indeed readily justifiable, to cluster sequences into representative subsets if called for by the subject of interest.*

due to the removal of degeneracies of 2 and 3 of the groups `seq.{0',2'}` (`sid` ***d3nira_***, ***d1ejga_***) and `seq.{1',8',9'}` (`sid` ***d5d8va_***, ***d3a38a_***, ***d3a39a_***) respectively, and the remaining 2 reductions due to the lowering of merge heights for the group `seq.{5',3',4'}` (`sid` ***d2dsxa_***, ***d5nw3a_***, ***d1yk4a_***) relative to other groups. The reductions would allow six groups, instead of only one presently, to be selected from the branches colored in green. In other words, the 11 clusters in Figure 3(b) for $\hat{\mathrm{E}}\mathrm{val}_2$ are more diverse than those in Figure 3(a) for $\mathrm{Eval}_2$ with sequence degeneracy and double- and triple-representations.

Similar to Case Study #2, the rates of PFN's and triangle inequality violations are high, at 97.5% and 92.4% respectively.

Beyond the first 180 domains above, in the ASTRAL compendium file `astral-scopedom-seqres-sel-gs-e100m-verbose-e-50-2.08.txt` there are a total of 302566 of domains factored into 58375 clusters at the threshold `Eval` of 1.0e−50, of which 44440 clusters are singletons (e.g. `seq.{0'..9'}` in Figure 3). Of these, 23515 are non-degenerate clusters each of a unique sequence, with the remaining 20925 in 4759 groups of degeneracy of up to 175 (e.g. groups `seq.{1',8',9'}` and `seq.{0',2'}` in Figure 3). Re-clustering with $\hat{\mathrm{E}}\mathrm{val}_2$ would have at least deselected 16166 domains with degenerate, redundant sequences (20925 − 4759), in favor of other domains and clusters of non-redundant sequences to be drawn from the 13935 complex, non-singleton clusters, significantly improving the sequence diversity within the entire subset.

**D. Case Study #4, distance function and metrication of protein sequence space**

Of the four axiomatic properties of a metric space (Methods, subsection G), positivity `M.13`, and symmetry `M.15`, are satisfied by both dissimilarity measures $\mathtt{Eval}_2$ and $\hat{\mathtt{E}}\mathtt{val}_2$:

| | | |
|---|---|---|
| positivity: | $\mathtt{Eval}_2$ | *cf.* `M.2`, `M.4` |
| | $\hat{\mathtt{E}}\mathtt{val}_2$ | *cf.* `R.1a`, `R.1` |
| symmetry: | $\mathtt{Eval}_2$ | *cf.* `R.3` |
| | $\hat{\mathtt{E}}\mathtt{val}_2$ | *cf.* `R.3a` |

Significantly, the reflexivity property, `M.14`, for self-distance of 0.0, is satisfied by ‘regularized’ dissimilarity $\hat{\mathtt{E}}\mathtt{val}_2$ only:

| | | |
|---|---|---|
| reflexivity: | $\hat{\mathtt{E}}\mathtt{val}_2$ | *cf.* `R.2a`, `R.2` |

whereas self-matching dissimilarity $\mathtt{Eval}_2$ is greater than 0: `M.4` and `M.2` with $\overline{\mathtt{s}} = \overline{\mathtt{q}}$, and illustrated in Table II, rows 5-7.

The remaining axiomatic property of metric space, triangle inequality, M.16, holds for the 'regularized' dissimilarity $\hat{\mathrm{E}}\mathrm{val}_2$ as demonstrated for sequences in Case Study #1, in Table II, rows 8,9. Formally, the relationship M.10 is rearranged algebraically to M.12 in which contributions from sequence lengths and alignment scores are refactored into two separate, bracketed terms. Rather than a formal proof of the deconstructed relationship M.12, either analytically or by enumeration and complete induction, $\mathrm{Eval}_2$, $\hat{\mathrm{E}}\mathrm{val}_2$, and the two bracketed terms in the expression are calculated from triplets of structure domains randomly taken from the SCOP database[6] for analysis. While $\mathrm{Eval}_2$ fails the condition of triangle inequality, i.e. the left-hand side of R.4, generalized to any sequence triplets, falls below y=0 in Figure 4(a), $\hat{\mathrm{E}}\mathrm{val}_2$ values on the other hand largely satisfy the condition of triangle inequality, i.e. the left-hand side of generalized R.4a is greater than 0 and falls to the right of x=0. Of the two bracketed terms in M.12 for $\hat{\mathrm{E}}\mathrm{val}_2$, the residual sequence length dependency term makes a smaller contribution than the second term of alignment scores (Figure 4(b)). As shown in red in the inset, for the small number of cases (4.7%) where the triangle inequality fails: (i) the violation is minimal, i.e. the left-hand side of M.12 has a small negative value, only marginally less than zero, and (ii) the alignment scores sum to 0 (the second bracketed term in M.12), with a relatively small, non-zero value for the sequence length term (the first bracketed term in M.12), which is traceable to two

Figure 4

---

**6** *For some structure domains in* SCOP *database, amino acid residues are given non-standard one-letter codes, such as* x, b *or* z, *if they are undetermined (*x*) or ambiguous (*b *or* z*) in X-ray diffraction experiments. Also letter* X *marks interruptions and discontinuities in the amino acid sequence of a structure domain. Although these non-standard amino acid letter codes are in the alphabet of substitution matrix for the calculation of alignment score (*Align *class of* Bio.Align *subpackage), sequences containing non-standard amino acids are outside the set $\mathcal{S}$ of normal, naturally occurring protein sequences (Methods, subsection G). These atypical structure domains are therefore by-passed in the random sampling of* SCOP *domains for data points in Figure 4.*

specific circumstances of the sequence lengths: either $t \neq u \neq v$, or $u = v$. These observations suggest that the interplay between sequence length and alignment score in the distance function $\hat{\mathtt{E}}\mathtt{val}_2$ (M.2, M.6, M.7) and in the evaluation of M.12 may play a significant role in the minimal violation of triangle inequality shown in Figure 4(b). Its origin notwithstanding, and however minor it may be, the violation nonetheless implies a lesser metric space (e.g. M.17 and M.18, but ***not*** M.16), for which $\hat{\mathtt{E}}\mathtt{val}_2$ is the distance function. To determine the weaker triangle inequality for the lesser metric space, consider first the data points to the right of the dividing line in Figure 4(b) with the relationship

$$\hat{\mathtt{E}}\mathtt{val}_2(\bar{v},\bar{t}) + \hat{\mathtt{E}}\mathtt{val}_2(\bar{t},\bar{u}) - \hat{\mathtt{E}}\mathtt{val}_2(\bar{v},\bar{u}) \geq 0$$

which is equivalent to

$$K' \cdot (\hat{\mathtt{E}}\mathtt{val}_2(\bar{v},\bar{t}) + \hat{\mathtt{E}}\mathtt{val}_2(\bar{t},\bar{u})) - \hat{\mathtt{E}}\mathtt{val}_2(\bar{v},\bar{u}) \geq 0 \qquad \text{R.5}$$

where $K' = 1$. Secondly, for data points to the left of the dividing line,

$$\hat{\mathtt{E}}\mathtt{val}_2(\bar{v},\bar{t}) + \hat{\mathtt{E}}\mathtt{val}_2(\bar{t},\bar{u}) - \hat{\mathtt{E}}\mathtt{val}_2(\bar{v},\bar{u}) = -\delta$$

where $\delta > 0$, which, upon rearrangement, becomes

$$K'' \cdot (\hat{\mathtt{E}}\mathtt{val}_2(\bar{v},\bar{t}) + \hat{\mathtt{E}}\mathtt{val}_2(\bar{t},\bar{u})) - \hat{\mathtt{E}}\mathtt{val}_2(\bar{v},\bar{u}) = 0 \qquad \text{R.6}$$

where $K''(\bar{t},\bar{u},\bar{v}) = 1 + \delta / (\hat{\mathtt{E}}\mathtt{val}_2(\bar{v},\bar{t}) + \hat{\mathtt{E}}\mathtt{val}_2(\bar{t},\bar{u})) > 1$, and $K'' > K'$. Let $\mathtt{Kappa} = \max(K''(\bar{t},\bar{u},\bar{v}))$ for all triplets $\{t, u, v\}$, then effectively combining both R.5 and R.6 above, for all data points in Figure 4(b),

$$\text{Kappa} \cdot \left( \hat{\text{E}}\text{val}_2(\bar{v},\bar{t}) + \hat{\text{E}}\text{val}_2(\bar{t},\bar{u}) \right) - \hat{\text{E}}\text{val}_2(\bar{v},\bar{u}) \geq 0$$

a relationship specified exactly for Kappa-relaxed triangle inequality M.17. K" is readily calculated from data in Figure 4(b), with a maximum of at least 1.0017 for domains ***d1lcnb_***, ***d1uiaa_*** and ***d4phia_*** among random triplets. Therefore, instead of triangle inequality M.16, $\hat{\text{E}}\text{val}_2$ satisfies a minimally relaxed triangle inequality M.17 with Kappa of about 1.0017[7], thus encoding a ***semi*** metric space for protein sequences in the SCOP database. The distribution of positives and negatives of the sum of the two bracketed terms of M.12, **viz**. those above or below the dividing line in Figure 4(b), remains invariant for different parameter values; for example, for open_gap_score of −13.0 ($\lambda$ and K of 0.292 and 0.071 respectively), Kappa is 1.0014.

It is worth noting that for the small and trivial set of first three sequences in Table I(a), the triangle inequality holds for the minimum space $\mathcal{S}$(seq.{1,b,z})⊕Eval without the need for the 'regularized' distance function $\hat{\text{E}}\text{val}$. Whether or not there may exist additional sequences that could expand the small space, it is reasonable to conjecture that most if not all sets of protein sequences of bioinformatic interests would be more similar to those from SCOP domain sequences than seq.{1,b,z}, for which triangle inequality be violated and database-wide operations susceptible to potential false negatives without regularized E value function $\hat{\text{E}}\text{val}$. Also worth noting is that the space $\mathcal{S}$(SCOP)⊕Eval can not be accommodated by any relaxed triangle inequality relationship M.17 of a reasonable value of Kappa due to the dominant and massive violations noted in Figure 4(b).

[7] *The value of* Kappa *is subject to an exhaustive enumeration of all triplets of* SCOP *domain sequences. However, from a close examination of the sequences of the domains* ***d1lcnb_***, ***d1uiaa_*** *and* ***d4phia_*** *and other triplets in Figure 4(b) violating the triangle inequality, specific sequence patterns were observed and the length dependency can be used for extrapolating the value of* Kappa *to a limit of 1.0436 (1.0398 for* open_gap_score *of -13.00), a theoretical upper bound for* ***any*** *set of polypeptide sequences. It would be interesting also to further trace the source of the small but evidently not insignificant deviation in the non-unitary* Kappa*, e.g. the length* n *in* M.2*, or various parameters in the calculation of alignment score* S*, factors that might strengthen the lesser metric space of a weakened triangle inequality* M.17 *to* M.16*, restoring* Kappa *back to* 1.

With the 'regularized' dissimilarity as the distance function, the metricity of protein sequence space may benefit from certain efficient search operations [17,21] to be exploited by similarity and other searches such as multiple sequence alignment for example [22]. In a recent study [19], a new E value for BLAST searches has been introduced with improved significance estimates (as well as certain computational and operational efficiencies). The new procedure improves the statistical significance as the primary goal, absent of any algebraic transformations similar to M.5, M.6 that are essential in the E value regularization. The query-length induced pattern of alignment score and E values shown respectively in columns 2 and 3 of Table II should be readily verifiable for the new E value calculations, to which the regularization procedure described herein would likely be applicable for sequence alignment comparisons over multiple queries. Moreover, the regularized dissimilarity scores based on the statistically improved E value [19] may be investigated particularly with regard to the Kappa value, for either a strict or a relaxed metricity (*cf.* Footnote *7*).

Beyond the simulated E value calculation reported herein, regularizing E value toward a dissimilarity measure and distance function for metricating protein sequence space may be further explored for the following. (i) In contrast to fixed parameter set (as noted in Methods, subsection B), for mixed parameters (e.g. BLOSUM matrices and/or gap scores, theoretically speaking), E value regularization (Table II) would be expected to be more nuanced, e.g. more complicated variants of expression M.12, in a manner likely dependent on specific mixed parameter sets, which would require closer examinations as necessary in practice. (ii) The relevant procedures in principle may be incorporated into the BLAST+ package (`http://blast.ncbi.nlm.nih.gov/Blast.cgi`) for example. Also, in the search for new protein families [16], a 'sequence landscape' was constructed from the results of 'all-against-all' sequence searches from `MMseqs2` [23], with the alignment coverage as threshold and E value as dissimilarity criterion. It would be interesting to examine E values in `MMseqs2` in the manner outlined in this report.

# Methods

All computations and analyses are carried out in `Python` (version 3.11), with the `Biopython` suite for bioinformatics [24], specifically the `Align` module in the suite for sequence alignment, and the `Bio.SCOP` subpackage [25] for accessing protein structure data in the hierarchical database `SCOP` [10,11], version 2.08 updated on 2023-01-06 [26]. `Python` libraries `SciPy` and `matplotlib` are used for clustering and dendrogram generation and for data plotting respectively. The system for simulated E value calculations is described in subsection B below.

For this study, the relevant functions are defined as follows:

A. Alignment score `S`, from the `PairwiseAligner` class in the `Align` package of `Biopython`,

$$S(\overline{s},\overline{q}) := \mathtt{Align.PairwiseAligner.score}(\overline{s},\overline{q}) \qquad \text{M.1}$$

where $\overline{s}$ and $\overline{q}$ are the amino acid sequences of subject $s$ and query $q$ respectively, with the alignment parameters for the `score` method: `open_gap_score`, −11.0, `extend_gap_score`, −1.0, `substitution_matrix`, 'BLOSUM62', and `Align.PairwiseAligner.mode`, 'global'.

B. E value, computed as a function of `S` [2,5], or bit-score `S'` [3,20],

$$\mathrm{Eval}(\overline{s},\overline{q}) := K\cdot m\cdot n\cdot e^{(-\lambda\cdot S(\overline{s},\overline{q}))} \qquad \text{M.2}$$

$$:= m\cdot n/2^{S'(\overline{s},\overline{q})} \qquad \text{M.2a}$$

where $m$ and $n$ are respectively the database size and the query length, with $m$ of $10^8$ [14], and $S'$ the bit-score [3,20],

$$S'(\overline{s},\overline{q}) := (\lambda \cdot S(\overline{s},\overline{q}) - \ln(K))/\ln(2) \qquad \text{M.3}$$

expressed in terms of alignment score $S$ (from M.1), and statistical parameters $\lambda$ and $K$ of 0.267 and 0.041 respectively [2,5-7], specific for the alignment parameters in Expr M.1. Since

$\mathrm{Eval}(\overline{s},\overline{q}) \neq \mathrm{Eval}(\overline{q},\overline{s})$ in general, the smaller of the two is assigned to $\mathrm{Eval}_2$ for the subject/query pair $s$ and $q$ [14]:

$$\mathrm{Eval}_2(\overline{s},\overline{q}) := \min(\mathrm{Eval}(\overline{s},\overline{q}),\mathrm{Eval}(\overline{q},\overline{s})) \qquad \text{M.4}$$

**N.B.** A computational study in protein sequence comparison study (NCBI blast, MMseqs2) often proceeds, at least at the start, with a fixed set parameters (such as dataset size, amino acid substitution matrix and gap scores and corresponding $\lambda$ and $K$ listed above), with which a baseline too may be established for the regularization of E value for dissimilarity measure. Of these parameters, the database size $m$ is a multiplicative factor of Eval (M.2, M.2a) and thus has no effect on the order of numerical values in column 3 of Table II. Effects of alternative sets of remaining parameters are further listed and noted in Table II.

C. Base E value,

$$\mathrm{Eval}_b(\overline{s},\overline{q}) := \sqrt{(\mathrm{Eval}(\overline{s},\overline{s}) \cdot \mathrm{Eval}(\overline{q},\overline{q}))} \qquad \textbf{M.5}$$

geometric mean of self-matching E values for subject and query $s$ and $q$, as a 'standardization' factor for Eval.

D. Distance E value, $\hat{\mathrm{E}}\mathrm{val}$ and $\hat{\mathrm{E}}\mathrm{val}_2$

$$\mathtt{\hat{E}val}(\overline{\mathtt{s}},\overline{\mathtt{q}}) := \ln(\mathtt{Eval}(\overline{\mathtt{s}},\overline{\mathtt{q}}) / \mathtt{Eval}_{\mathtt{b}}(\overline{\mathtt{s}},\overline{\mathtt{q}})) \qquad \mathbf{M.6}$$

$$\mathtt{\hat{E}val}_2(\overline{\mathtt{s}},\overline{\mathtt{q}}) := \min(\mathtt{\hat{E}val}(\overline{\mathtt{s}},\overline{\mathtt{q}}),\mathtt{\hat{E}val}(\overline{\mathtt{q}},\overline{\mathtt{s}})) \qquad \mathbf{M.7}$$

E. <u>‘Regularized’ similarity</u> $\hat{\mathtt{S}}$ and $\hat{\mathtt{S}}_2$

$$\hat{\mathtt{S}}(\overline{\mathtt{s}},\overline{\mathtt{q}}) := e\hat{}(-\mathtt{\hat{E}val}(\overline{\mathtt{s}},\overline{\mathtt{q}})) \qquad \mathtt{M.8}$$

$$\hat{\mathtt{S}}_2(\overline{\mathtt{s}},\overline{\mathtt{q}}) := e\hat{}(-\mathtt{\hat{E}val}_2(\overline{\mathtt{s}},\overline{\mathtt{q}})) \qquad \mathtt{M.9}$$

following general exponential relationship between similarity and dissimilarity measures [8,9].

F. <u>Triangle inequality</u> (*cf.* $\mathtt{M.16}$ below) for the sequence triplet $\{\mathtt{t,u,v}\}$ states that:

$$\mathtt{\hat{E}val}_2(\overline{\mathtt{v}},\overline{\mathtt{t}}) + \mathtt{\hat{E}val}_2(\overline{\mathtt{t}},\overline{\mathtt{u}}) - \mathtt{\hat{E}val}_2(\overline{\mathtt{v}},\overline{\mathtt{u}}) \geq 0 \qquad \mathtt{M.10}$$

where, without loss of generality, $\mathtt{\hat{E}val}_2(\overline{\mathtt{v}},\overline{\mathtt{u}})$ is the largest among the three pairwise $\mathtt{\hat{E}val}_2$'s.

Substituting $\mathtt{M.7}$ for $\mathtt{\hat{E}val}_2$, and then $\mathtt{M.6}$ and $\mathtt{M.5}$ for $\mathtt{\hat{E}val}$, and $\mathtt{M.2}$ for $\mathtt{Eval}$, the first term on the left-hand side of $\mathtt{M.10}$, $\mathtt{\hat{E}val}_2(\overline{\mathtt{v}},\overline{\mathtt{t}})$, becomes:

$$\frac{1}{2} \cdot \min(\ln(t/v), \ln(v/t)) -$$

$$\lambda \cdot (\mathtt{S}(\overline{\mathtt{v}},\overline{\mathtt{t}}) - \frac{1}{2} \cdot (\mathtt{S}(\overline{\mathtt{v}},\overline{\mathtt{v}}) + \mathtt{S}(\overline{\mathtt{t}},\overline{\mathtt{t}}))) \qquad \mathtt{M.11}$$

where $\overline{\mathtt{t}}$ and $\overline{\mathtt{v}}$ are the sequences of $\mathtt{seq.t}$ and $\mathtt{seq.v}$ and $t$ and $v$ are their lengths. Substituting the expanded form of $\mathtt{\hat{E}val}_2(\overline{\mathtt{v}},\overline{\mathtt{t}})$ above ($\mathtt{M.11}$), and similarly for the two remaining terms, $\mathtt{M.10}$ algebraically becomes the following, with the left-hand side being the linear sum of two bracketed terms:

$$\frac{1}{2} \cdot \left[\ \min\left(\ln(t/v), \ln(v/t)\right) + \min\left(\ln(t/u), \ln(u/t)\right) - \min\left(\ln(u/v), \ln(v/u)\right)\right] +$$

$$\lambda \cdot \left[S(\overline{t},\overline{t}) + S(\overline{v},\overline{u}) - S(\overline{v},\overline{t}) - S(\overline{t},\overline{u})\right] \geq 0 \qquad \textbf{M.12}$$

In this deconstructed form, contributions from sequence lengths and from alignment scores to the metric property (subsection G) of triangle inequality M.10 are refactored as two individual terms that are analyzed more readily (Results and Discussion).

G. Metric space, is a space, denoted as $\mathcal{S}\oplus d$ where formally a set $\mathcal{S}$ is defined together with function $d(x,y)$, or metric, for a distance measure between set members, i.e. protein sequences $x$ and $y$ in $\mathcal{S}$, with the following four axiomatic properties for all members $x,y$ [21]:

positivity

$$d(x,y) \geq 0 \quad \text{for set members } x,y \qquad \text{M.13}$$

reflexivity (identity of indiscernibles)

$$d(x,y) = 0 \quad \text{if and only if } x = y \qquad \text{M.14}$$

symmetry

$$d(x,y) = d(y,x) \qquad \text{M.15}$$

triangle inequality

$$d(x,y) \le d(x,z) + d(z,y) \qquad \text{M.16}$$

A 'lesser' metric space is a metric space with some of the axiomatic properties modified to a more relaxed, weakened condition. Specifically, a ***semi*** metric space is a space for which triangle inequality M.16 is replaced with a weaker inequality:

Kappa-relaxed triangle inequality [27]

$$d(x,y) \le Kappa \cdot (d(x,z) + d(z,y)) \qquad \textbf{M.17}$$

where $Kappa \ge 1$, or

quadrilateral inequality [28]

$$d(x,y) \le d(x,z) + d(z,w) + d(w,y) \qquad \text{M.18}$$

These two ***semi*** metric spaces are also known as the ***b***-metric and the ***g***-metric space respectively.

H. Systems information. Data processing and computations were carried out on a Linux virtual machine (Debian 12 operating system) hosted in Qubes OS hypervisor (4.2.3) running on a Dell 3505 computer, with dual-core Ryzen 5 processor and 16 megabytes of memory.

## Key Points

- The E value of BLAST similarity searches is qualified as 'bits' dissimilarity.
- Leveraging the E value framework, we formulate a ‘regularized’ dissimilarity that resolves certain peculiarities in the assessment of sequence similarity (Case Study #1).
- The ‘regularized’ dissimilarity satisfies the important property of zero self-distance for a proper distance function, and transparently addresses identity as a limiting and ultimate form of similarity (Case Study #2).
- Validated for triangle inequality, the fourth and final axiomatic property of a metric space (Case Study #4), the ‘regularized’ dissimilarity is a proper distance function for metricating protein sequence space.
- Put in practice, the ‘regularized’ dissimilarity improves the sequence diversity of clustering and subset selection and helps identify potential false negatives in near neighbor searches (Case Study #3).

## Acknowledgments

This is an independent research work, with no academic, governmental, corporate, or institutional funding or sponsorship. The author thanks the reviewers for their comments and critiques which greatly helped improve this report.

# Figure Legends

## Figure 1

'Regularization' of E value for dissimilarity measure. (a) $\texttt{Eval}$ and $\texttt{Eval}_{\texttt{b}}$ (scale on the left), and $\hat{\texttt{E}}\texttt{val}$ and $\triangle\hat{\texttt{E}}\texttt{val}$ (scale on the right) for pairs of sequences in Table I(a). The four curves are (from the top): $\texttt{Eval}$ (in light green), $\texttt{Eval}_{\texttt{b}}$ (purple), $\hat{\texttt{E}}\texttt{val}$ (blue), and $\triangle\hat{\texttt{E}}\texttt{val}$ (green), also identified by color-coded labels on Y-axis. $\triangle\hat{\texttt{E}}\texttt{val}$ is the absolute value of $\hat{\texttt{E}}\texttt{val}(\overline{\texttt{s}},\overline{\texttt{q}}) - \hat{\texttt{E}}\texttt{val}(\overline{\texttt{q}},\overline{\texttt{s}})$ displayed in 20-fold. $\texttt{Eval}(\overline{\texttt{az}},\overline{\texttt{a}})$ is larger than $\texttt{Eval}(\overline{\texttt{b}},\overline{\texttt{1}})$ (in light green, first curve from top, and also shown as merge heights in (b)), and $\hat{\texttt{E}}\texttt{val}(\overline{\texttt{az}},\overline{\texttt{a}})$ is smaller than $\hat{\texttt{E}}\texttt{val}(\overline{\texttt{b}},\overline{\texttt{1}})$ (in blue, third curve from top, and also in (c)). Since $\hat{\texttt{E}}\texttt{val}$ does not alter the order of the first three points of the $\texttt{Eval}$–$\hat{\texttt{E}}\texttt{val}$ curve (in blue, numerical data in Table II), the partial curve from these points are monotonically increasing for the minimum space $\mathcal{S}(\texttt{seq.\{1,b,z\}})\oplus\texttt{Eval}$. The break of monotonicity of the $\texttt{Eval}$–$\hat{\texttt{E}}\texttt{val}$ curve, notably at the $\texttt{seq.az/a}$ pair, signifies a potential false negative of the 'bits' dissimilarity measure $\texttt{Eval}$. (b) Clustering of the sequences in Table I(a) on $\texttt{Eval}$. Note that $\texttt{Eval}(\overline{\texttt{az}},\overline{\texttt{a}}) > \texttt{Eval}(\overline{\texttt{z}},\overline{\texttt{1}})$. (c) Clustering on $\hat{\texttt{E}}\texttt{val}$. Here $\hat{\texttt{E}}\texttt{val}(\overline{\texttt{az}},\overline{\texttt{a}}) = \hat{\texttt{E}}\texttt{val}(\overline{\texttt{z}},\overline{\texttt{1}})$. Given the small variations of $\hat{\texttt{E}}\texttt{val}$'s from one set of $\lambda$,$\texttt{K}$ parameters to another (Table II, column 5), the clustering dendrogram remains essentially invariant with little change in the ordering of merging heights (*cf.* Footnote ***4***).

## Figure 2

Clustering of 40 domains (dom.0 through dom.39) in Case Study #2: (a) Clustering on Eval. (b) Clustering on Êval. Three groups are of degeneracy of 2: dom.{1,2} and dom.{3,4}, in branch E and C respectively, and dom.{18,20} in branch B. One group of degeneracy of 3: dom.{34,35,39} in branch B. Groups of degeneracy of 4: dom.{11,12,17,19} in branch D, and two others in branch B. One group of degeneracy of 8: dom.{22,25,27,28,29,30,31,32} in branch B.

## Figure 3

Clustering of 180 domains (dom.0' through dom.179') in Case Study #3: (a) Clustering on Eval. A threshold at 1e−50 produces the same 11 clusters in the SCOP subset file at the same E value threshold of 1e−50 in the ASTRAL compendium. (b) Clustering on Êval. The threshold of 55.0020 generates 11 clusters, the same number of clusters as in (a).

## Figure 4

Distributions of numerical values for the metric property of triangle inequality. (a) Comparison of the left-hand side of M.10 for Eval$_2$ and Êval$_2$, demonstrating triangle inequality in general does not hold for the former. (b) Relationship of the two bracketed terms in Expr M.12, with the inset showing a close-up view of the area near the origin. The blue line, extending from the origin through the point (1,−1), is a dividing line that marks the separation of the area to its left, x+y<0, from the area to its right, x+y>0 where the triangle inequality holds. There are only a relatively small number of cases (4.7%) in which triangle inequality is violated minimally (along the Y-axis and shown in red in inset).

NB. For the space $\mathcal{S}$(`SCOP`)⊕`Eval`, the triangle inequality is violated in nearly all cases (99.1%), with predominantly high K" values as large as e+55.

Figure 1

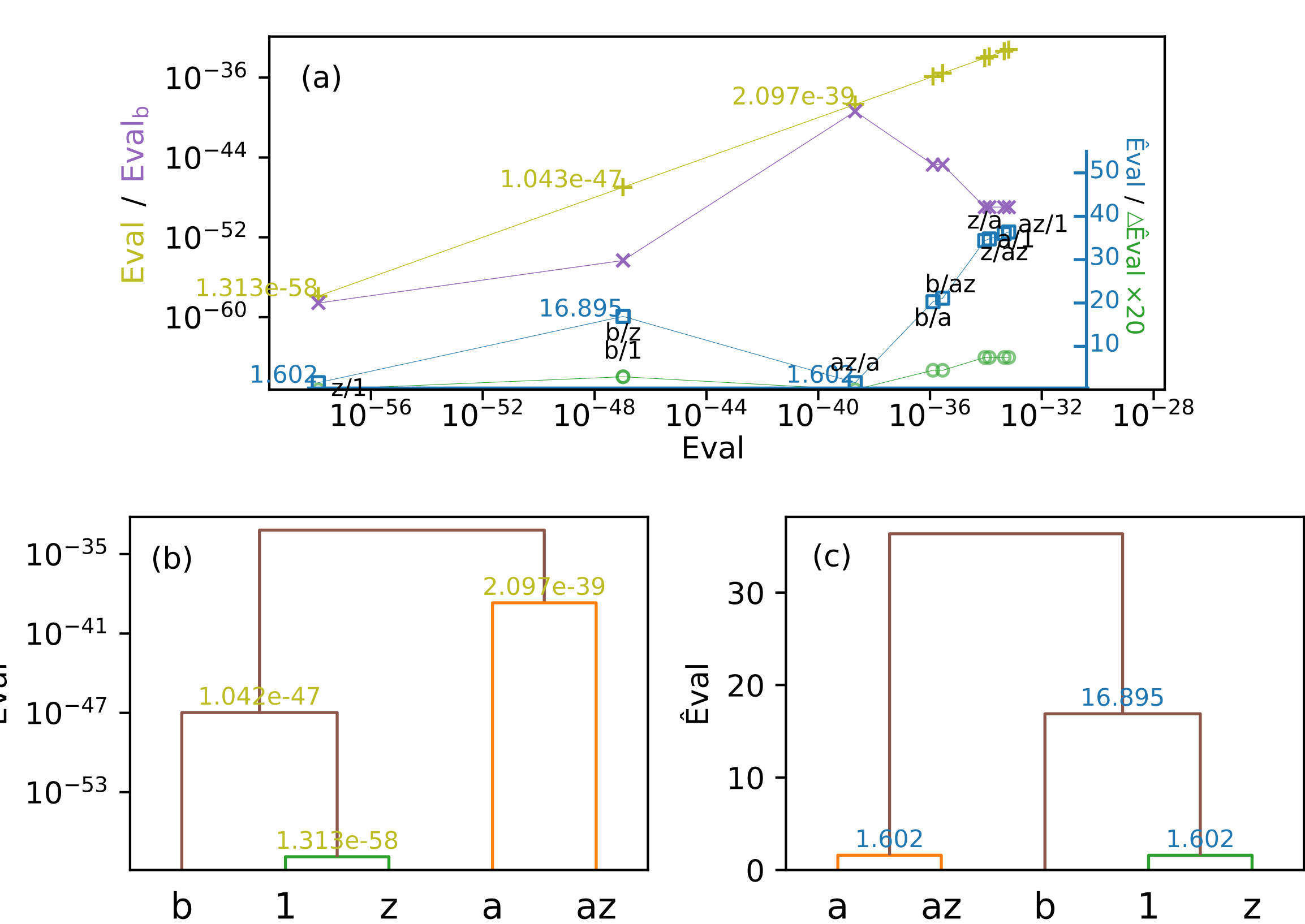

# Figure 2

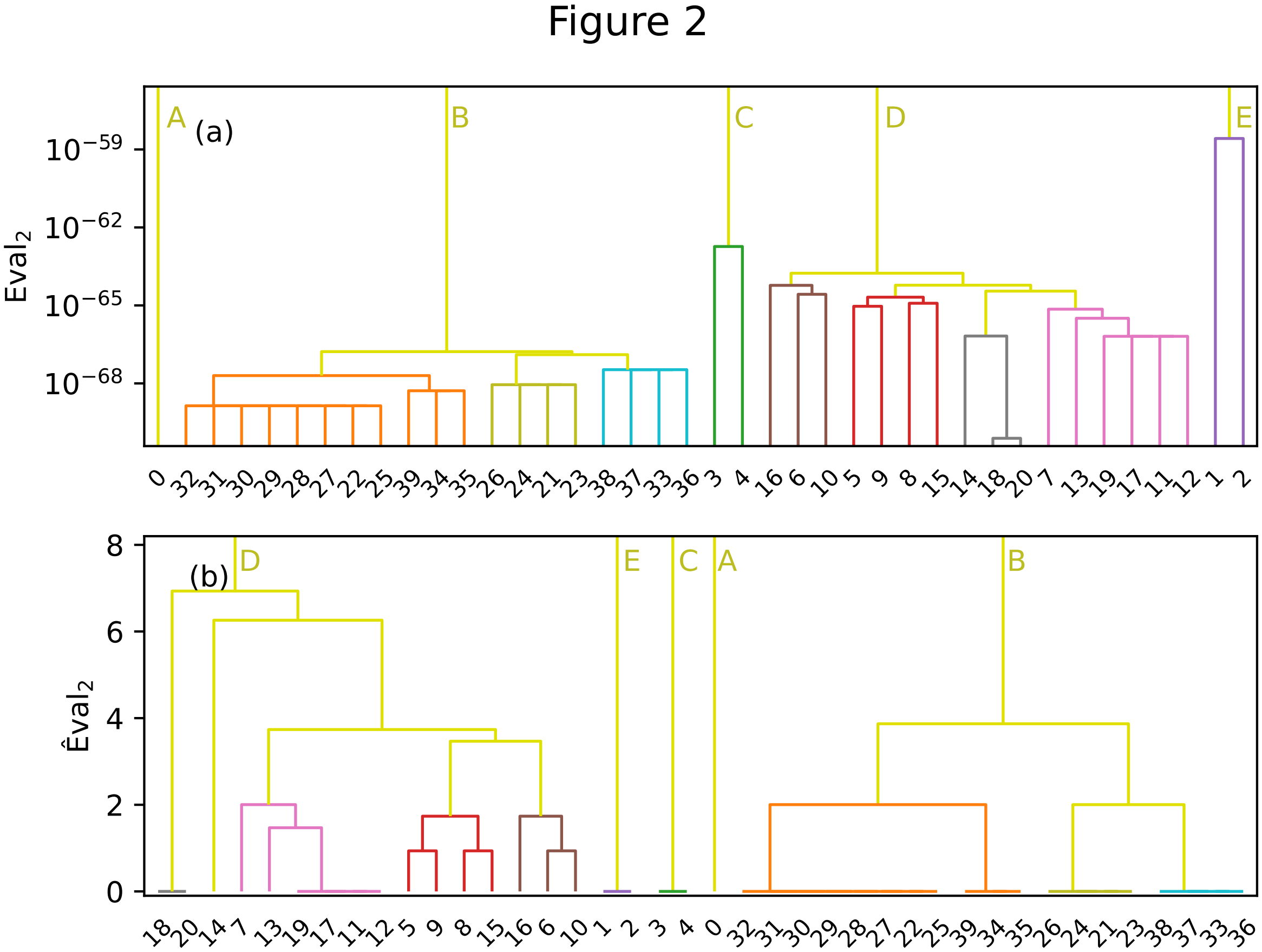

# Figure 3

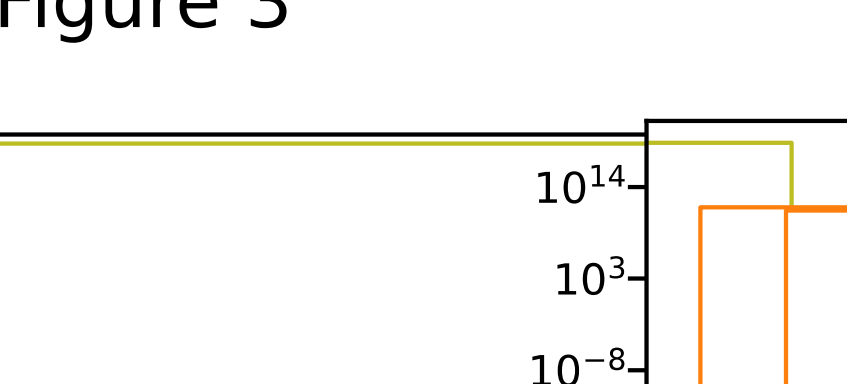

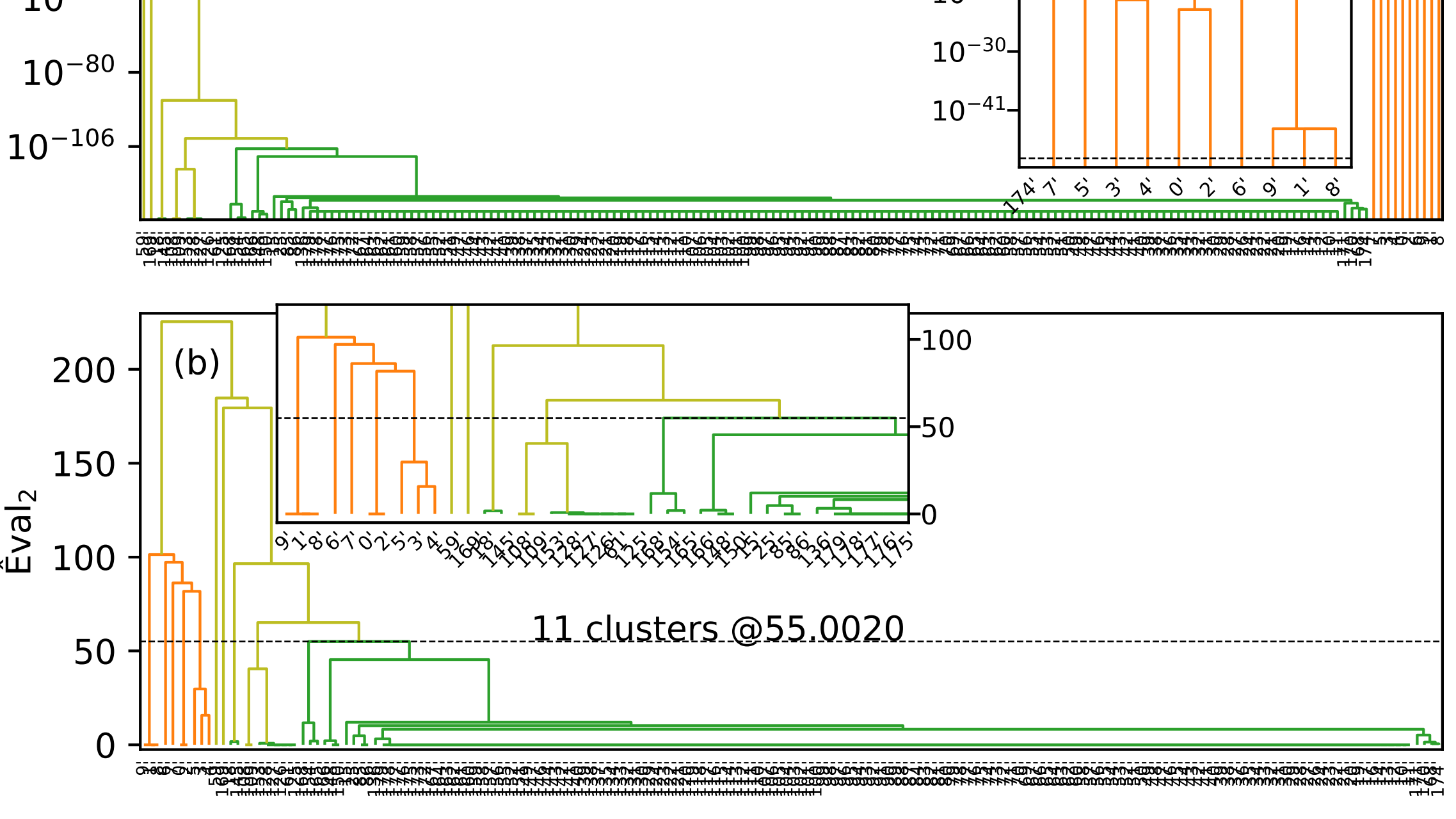

Figure 4

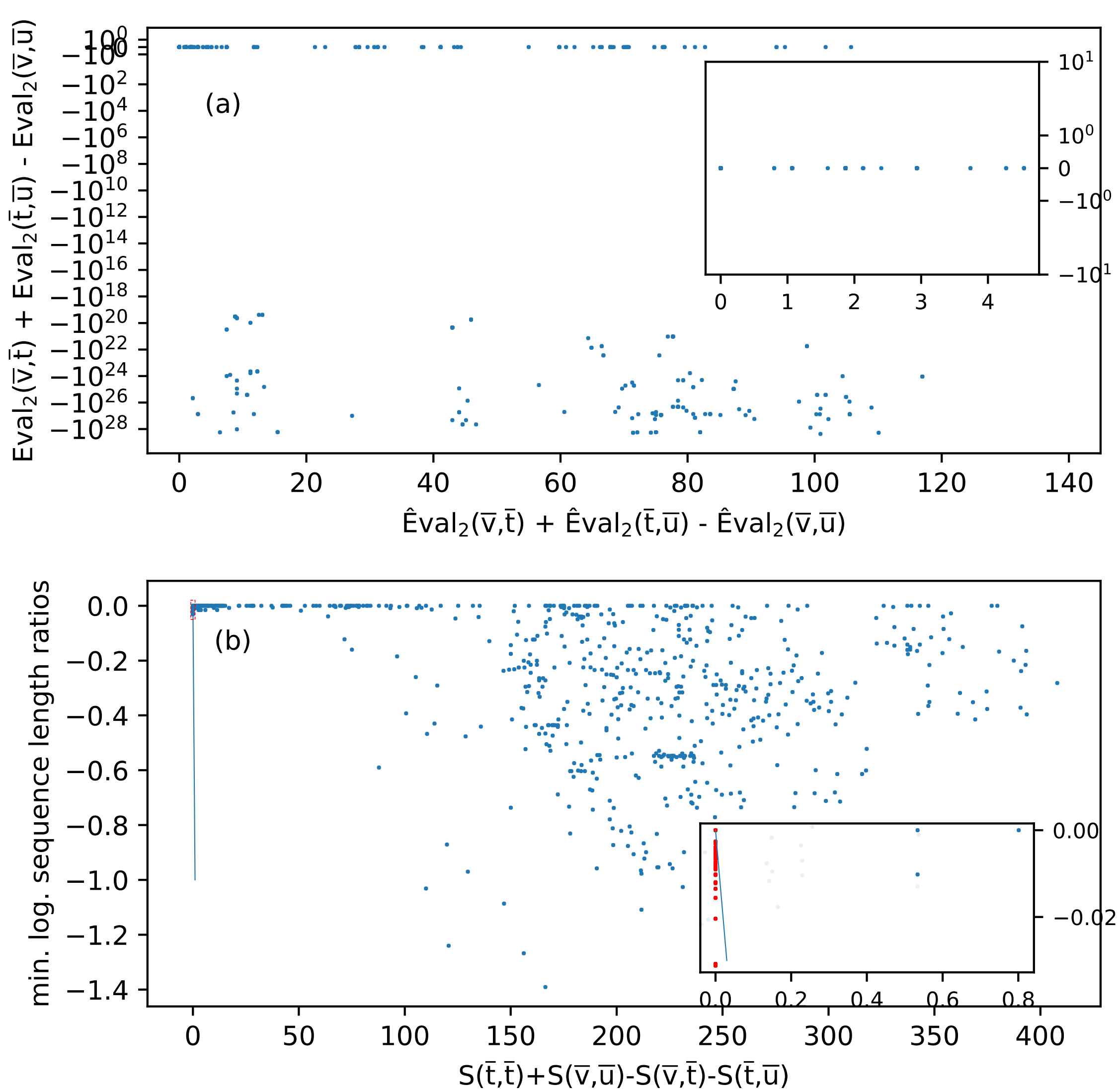

Table I. Amino acid sequences and their similarity and dissimilarity scores for Case study #1

(a)

| sequence name, s | amino acid sequence, $\bar{s}$ | notes |
|---|---|---|
| 1* | slfeqlggqaavqavtaqfyaniqadatvatffngidmpnqtnktaaflc<br>aalggpnawtgrnlkevhanmgvsnaqfttvighlrsaltgagvaaalve<br>qtvavaetvrgdvvtv | SCOP domain ***d1dlwa_*** (length 116 residues) |
| z | slfeqlggqaavqavtaqfyaniqadatvatffngidmpnqtnktaaflc<br>aalggpnawtgrnlkevhanmgvsnaqfttvighlrsaltgagvaaalve<br>qtvavaetvrgdvv**q**v | seq.1 with t-to-**q** substitution at position 115 |
| b | slfeqlggqaavqavtaqfyaniqadatvatffngidmpnqtnktaaflc<br>aalggpnawtgrnlkevhanmgvsnaqfttvighlrsaltgagvaaalve | residues 1-100 of seq.1 |
| az | slfeqlggqaavqavtaqfyaniqadatvatffngidmpnqtnktaaflc<br>aalggpnawtgrnlkevhanmgvsnaqft**q** | seq.a with t-to-**q** substitution at position 80 |
| a | slfeqlggqaavqavtaqfyaniqadatvatffngidmpnqtnktaaflc<br>aalggpnawtgrnlkevhanmgvsnaqftt | residues 1-80 of seq.1 |

(b)

| subject | query | alignment score (similarity) | E value (dissimilarity) |
|---|---|---|---|
| z | 1 | 574.0 | 1.313e-58 |
| b | 1 | 480.0 | 1.043e-47 |
| az | 1 | 361.0 | 6.560e-34 |
| az | a | 407.0 | 2.097e-39 |

(a) Amino acid sequence of SCOP domain ***d1dlwa_***, seq.1, and four derivative sequences for Case Study #1. seq.1 is fetched with Biopython subpackage Bio.SCOP (an interface to SCOP database), class and method Scop.getDomains() [1].
(b) Alignment scores and E values (calculated according to M.1 and M.2) of four subject,query pairs for Table II.

* Sequence of domain id 1 (dom.1) doubles as the sequence name.

Table II. Regularization of similarity and dissimilarity scores for sequences in Table I

| | s/q (subject,query) | $S(\overline{s},\overline{q})$ ('bits' similarity) | $\mathrm{Eval}(\overline{s},\overline{q})$ ('bits' dissimilarity) | $\mathrm{Eval}_b(\overline{s},\overline{q})$ | $\hat{\mathrm{E}}\mathrm{val}(\overline{s},\overline{q})$ ('regularized' dissim.) | $\hat{S}(\overline{s},\overline{q})$ ('regularized' sim.) |
|---|---|---|---|---|---|---|
| | z/1 | 574.0 | 1.313e-58 [1.332e-64] | 2.645e-59 | **1.602** [**1.752**] | **0.755** |
| | b/1 | 480.0 | 1.043e-47 [1.989e-52] | 4.793e-55 | 16.895(**16.747**) [**18.906**] | 1.722e-07 (**1.997e-07**) |
| | az/1 | 361.0 | 6.560e-34 [2.452e-37] | 1.057e-49 | 36.364(**35.993**) [**39.964**] | 1.611e-16 (**2.336e-16**) |
| | az/a | 407.0 | **2.097e-39** [**1.384e-43**] | 4.226e-40 | **1.602** [**1.752**] | **0.755** |
| self-matching | 1/1 | 580.0 | 2.645e-59 | 2.645e-59 | 0.0 | 1.0 |
| self-matching | b/b | 506.0 | 8.684e-51 | 8.684e-51 | 0.0 | 1.0 |
| self-matching | a/a | 413.0 | 4.225e-40 | 4.225e-40 | 0.0 | 1.0 |
| triangle inequality | z/a | | **9.116e-35** | | **34.391** [**38.212**] | |
| triangle inequality | z/az | | **4.524e-34** | | **35.993** [**39.964**] | |
| triangle inequality | a/1 | | 1.322e-34 | | 34.762(**34.391**) [**38.212**] | |
| triangle inequality | b/z | | 1.043e-47 | | 16.895(**16.747**) | |

Rows 1 to 4: Regularization of alignment score and E value (columns 2 and 3) to corresponding dissimilarity and similarity measures (columns 5 and 6). Values in bold face are the smaller of two dissimilarity/similarity scores when subject and query exchange positions. Rows 5-7: self-matching alignment scores and E values. Rows 8-9 (along with row 4), for checking triangle inequality of dissimilarity: 9.116e-35+2.097e-39≈9.116e-35<4.524e-34 for `Eval`, and 34.391+1.602=35.993 for `Êval`, and similarly for row 10 along with rows 3,4. Varying parameters in the calculation of alignment score introduces relatively small changes in numerical values but not the overall relationships: for example `Eval`'s and `Êval`'s for the gap penalty `open_gap_score` of −13.0 ($\lambda$ of 0.292 and $K$ of 0.071) are shown in square brackets in columns 3 and 5, and for substitution matrices `BLOSUM90` and `PAM250` (data not shown).

NB. (1) Unlike those among `seq.{1,z,az,a}` above, the triangle inequality holds for the triplet `seq.{1,b,z}`, rows 1,2, and 11, column 3, for which the `Eval`–`Êval` relationship is monotonic (the initial part of the curve in blue, third curve from the top and marked with squares, in Figure 1(a)). (2) For the `az/a` pair, the recovery of any potential false negative for `Eval` as a dissimilarity measure appears as a break in the monotonicity of the `Eval`–`Êval` curve.

Table III. Amino acid sequences of degenerate pairs and their similarity and dissimilarity scores

(a)

| domain id, d | amino acid sequence, $\bar{d}$ | notes |
|---|---|---|
| 1 | slfeqlggqaavqavtaqfyaniqadatvatffngidmpnqtnktaaflc<br>aalggpnawtgrnlkevhanmgvsnaqfttvighlrsaltgagvaaalve<br>qtvavaetvrgdvvtv | SCOP domain ***d1dlwa_*** (length 116 residues) |
| 2 | slfeqlggqaavqavtaqfyaniqadatvatffngidmpnqtnktaaflc<br>aalggpnawtgrnlkevhanmgvsnaqfttvighlrsaltgagvaaalve<br>qtvavaetvrgdvvtv | SCOP domain ***d1uvya_*** (length 116 residues) |
| 3 | slfaklggreaveaavdkfynkivadptvstyfsntdmkvqrskqfafla<br>yalggasewkgkdmrtahkdlvphlsdvhfqavarhlsdtltelgvpped<br>itdamavvastrtevlnmpqq | SCOP domain ***d1dlya_*** (length 121 residues) |
| 4 | slfaklggreaveaavdkfynkivadptvstyfsntdmkvqrskqfafla<br>yalggasewkgkdmrtahkdlvphlsdvhfqavarhlsdtltelgvpped<br>itdamavvastrtevlnmpqq | SCOP domain ***d1uvxa_*** (length 121 residues) |
| 18 | gllsrlrkrepisiydkiggheaievvvedfyvrvladdqlsaffsgtnm<br>srlkgkqveffaaalggpepytgapmkqvhqgrgitmhhfslvaghlada<br>ltaagvpsetiteilgviaplavdvtsgesttapv | SCOP domain ***d1s56b_*** (length 135 residues) |
| 20 | gllsrlrkrepisiydkiggheaievvvedfyvrvladdqlsaffsgtnm<br>srlkgkqveffaaalggpepytgapmkqvhqgrgitmhhfslvaghlada<br>ltaagvpsetiteilgviaplavdvtsgesttapv | SCOP domain ***d1s61b_*** (length 135 residues) |

(b)

| s/q (subject,query) | $S(\bar{s},\bar{q})$ ('bits' similarity) | $\mathrm{Eval}(\bar{s},\bar{q})$ ('bits' dissimilarity) | $\hat{\mathrm{E}}\mathrm{val}(\bar{s},\bar{q})$ ('regularized' dissim.) | $\hat{S}(\bar{s},\bar{q})$ ('regularized' sim.) |
|---|---|---|---|---|
| 1/2 | 580.0 | 2.645e-59 | 0.0 | 1 |
| 3/4 | 616.0 | 1.846e-63 | 0.0 | 1 |
| 18/20 | 680.0 | 7.809e-71 | 0.0 | 1 |

(a) Among the first 40 domains fetched from `SCOP` database (`Biopython` interface package `Bio.SCOP`, class and method `Scop.getDomains()[0:40]`) for Case Study #2, three domain pairs (domain names shown in bold italics in the Notes column) are doubly-degenerate, sharing pairwise identical sequences: `dom.{1,2}`, `dom.{3,4}`, and `dom.{18,20}`.
(b) Similarity and dissimilarity scores.